\documentclass[10pt,aps,prb,endfloats,superscriptaddress,raggedbottom]{revtex4}
\usepackage{graphicx,latexsym}
\usepackage{amsmath}

\begin{document}
\title
{\bf The Fano regime of one-dot Aharonov-Bohm
interferometers}

\author{Valeriu Moldoveanu}
\affiliation{National Institute of Materials Physics, P.O. Box MG-7,
Bucharest-Magurele, Romania}
\author{Mugurel \c{T}olea}
\affiliation{National Institute of Materials Physics, P.O. Box MG-7,
Bucharest-Magurele, Romania}
\author{Vidar Gudmundsson}
\affiliation{Science Institute, University of Iceland, Dunhaga 3, IS-107
Reykjavik, Iceland}
\author{Andrei Manolescu}
\affiliation{National Institute of Materials Physics, P.O. Box MG-7,
Bucharest-Magurele, Romania}
\affiliation{Science Institute, University of Iceland, Dunhaga 3, IS-107
Reykjavik, Iceland}

\begin{abstract}
We use the Landauer-B\"{u}ttiker formalism to study the mesoscopic Fano effect in
Aharonov-Bohm rings with an embedded two-dimensional noninteracting dot.
The magnetic field dependence of the dot levels leads to a global shift of
the Fano lines which becomes important for small ring/dot area ratios.
As the magnetic field is varied the Fano
dips move periodically from one side of the peak to the other,
as reported by Kobayashi {\it et al.} [Phys. Rev. Lett. {\bf 88}, 256806 (2002)]
We show that this effect appears due to a specific magnetic control of the difference
between the phase of the single nonresonant path via the free arm of the ring and
the global phase of all trajectories involving resonant tunnelings through the dot.

\end{abstract}
\maketitle
\section{Introduction}
The many similarities shared by mesoscopic devices based on quantum dots
(QD) and atomic or molecular systems justify experiments that can test
at mesoscopic level effects encountered earlier in molecular physics.
 An important illustration of the above idea is the recent
experiment carried out by Kobayashi {\it et al.} \cite{F1,F2}
that revealed the counterpart of the Fano effect from atomic physics \cite{Fano}
in a hybrid mesoscopic system, namely an Aharonov-Bohm  interferometer
with an embedded dot. The Fano effect is a resonance produced by
the quantum interference between a discrete state embedded in a continuous spectrum and
the continuous spectrum. In Ref. \cite{F1}
the discrete level is provided by the resonant transport through the QD,
and the continuous background comes from the free arm of the ring.

The mesoscopic Fano effect leads to a series of
asymmetric conductance peaks appearing as the gate voltage pushes the
dot levels across the Fermi energy of the leads. The experiment of
Kobayashi {\it et al.} presented even more interesting features
: i) The orientation of the Fano line shape changes periodically with the magnetic flux.
 ii) Between two Fano line shapes the conductance exhibits clear AB oscillations
proving the transport coherency. iii) The phase of the oscillations changes
by $\pi$ across a resonance.

The Fano effect in AB rings with quantum dots was predicted theoretically
by Bu{\l}ka {\it et al.} \cite{BS} The conductance of a ring with a single level
QD was computed in the Keldysh formalism by Hofstetter {\it et al.}
\cite{HKS} and the interplay between the Fano and Kondo effects has
been discussed.
Later on the Fano interference was discussed in the context of phase measurements
in Aharonov-Bohm interferometers \cite{AEY,AEI,EAIL}.
The calculations were done for a triangular tight-binding interferometer for
which the scattering matrix can be computed exactly. Like the quantum dot, the reference
arm of the ring was restricted to a single site.
A similar model was considered by Ueda {\it et al.} \cite{Ueda}
Kim and Hershfield \cite{KH} utilized a more complex description of the quantum dot
for studying the role of the sign in the tunneling matrix on the transmittance zeros and phase.

In the present work we attempt a theoretical description of the
Fano interference for a two-dimensional
dot and a much larger ring (in Ref. \cite{F1} the dot accommodates up to 80
electrons while the ratio between the area of the ring and
the dot area is around 56). There are two reasons for this generalization.
First, one can describe the effects of the magnetic field on the dot levels and
secondly, the investigation of finite size effects on the Fano interference observation
 is possible. Another aim of our study is to discuss the tuning of the
 Fano resonance by varying the magnetic field.
 The formalism we used to calculate the conductance has been described in
Ref. \cite{MTAT} and is summarized in Section II. It relies on the
Landauer-B\"{uttiker} formula and is similar to the scattering approach developed
by Hackenbroich and Weidenm\"{u}ller. \cite{HW2,Ha}

Since the Coulomb repulsion plays an important role in QD systems it is natural
to question about correlations effects. In the first experiment of
Kobayashi {\it et al.} \cite{F1} the Kondo effect was not detected but in a
recent work \cite{F3} a Fano-Kondo anti-resonance was observed in the case of a
quantum wire with a side-coupled QD, provided the lead-dot coupling is strong.
This issue was discussed theoretically in \cite{MSU,OAY}.
For weak coupling the correlation between the localized spins in the dot
and the incident ones from the leads is unlikely and one recovers
the Fano line shapes.

In the present work we consider only weak dot-ring couplings as in Ref.\cite{F1}
and we include neither the spin nor the electron-electron interaction.
Section III contains the numerical results and their
discussion. Section IV concludes the paper.

\section{The model}

We consider a mesoscopic ring connected to electron reservoirs by two
 semi-infinite leads. A noninteracting quantum dot is
embedded in the upper arm of the ring. The system is schematically shown
in Fig.\ 1, indicating as well the
labeling of the leads and the contact points. We use here a tight-binding representation
so the dot is modeled by a discrete lattice. The Aharonov-Bohm interferometer (I)
 is formed by coupling two sub-systems: a truncated ring (R) and the dot (D).
 Its Hamiltonian is then conveniently written in matrix form:
\begin{eqnarray}
H^I=
\begin{pmatrix}
      H^D & H^{DR}\cr
      H^{RD} & H^R\cr
\end{pmatrix},
\end{eqnarray}
where $H^{DR}$ and $H^{RD}$ are the ring-dot and dot-ring coupling terms:
\begin{equation}
H^{DR}+H^{RD}
=\tau\sum_{m=a,b}(e^{-i\varphi _m }|m\rangle\langle 0m|+h.c).
\end{equation}
The notation $0m$ stands for the site of the ring which is the closest
one to the dot.
$\tau$ gives the ring-dot coupling and
  $\varphi _m$ is the Peierls phase associated with the pair of sites
$|0m\rangle$, $|m\rangle$. Similar Peierls phases enter in both $H^D$ and $H^R$.
We do not give explicitly the tight-binding Hamiltonian $H^D$ of the dot.
We just stress that it contains an on-site constant $-eV_g$
simulating the plunger voltage used in experiments. We take $e=h=1$.

The conductance (i.e transmittance) of one-dot AB interferometer was obtained via the
 Landauer-B\"{u}ttiker formalism in Ref. \cite{MTAT}
Denoting by $\tau_L$ the lead-ring coupling and by $E_F$ the Fermi energy on leads
one gets the following
formula for $\alpha\neq\beta$ at $T=0K$
(see Ref. \cite{MTAT} for further details):
\begin{eqnarray}\label{gab}
g_{\alpha\beta}(E_F)=4\tau_L^4\sin^2k \left |
{\tilde G}^R_{\alpha\beta}+
\tau^2e^{i\theta_{mn}}
{\tilde G}^R_{\alpha m}{\tilde G}^D_{mn}{\tilde G}^R_{n\beta}\right |^2,
\end{eqnarray}
where $\theta_{mn}=\varphi_m-\varphi_{n}$, summations over $m,n$ are understood
and we introduced two effective Green functions describing {\it individually} the dot
and the truncated ring:
\begin{eqnarray}\label{GReff}
{\tilde G}^R_{\alpha m}(z)&:=& \langle \alpha|(H^R-\Sigma^L(z)-z)^{-1}
|0m\rangle\\\label{H_eff}
{\tilde G}^D_{mn}(z)&:=&\langle m|(H^D-\Sigma^D(z)-z)^{-1}|n\rangle.
\end{eqnarray}
Here $\Sigma^L(z)$ is the leads' self-energy and
$\Sigma^D(z):=H^{DR}(H^R-\Sigma^L-z)^{-1}H^{RD}$ is the self-energy due to the coupling to the ring. Formula (\ref{gab}) allows us to numerically compute $g_{\alpha\beta}$.
The first term represents the free transport via the lower arm and the second term
embodies {\it all} the trajectories that involve at least one resonant tunneling.
Thus Eq. (\ref{gab}) is more complicated than what one gets in a two-slit experiment.
The mesoscopic Fano effect differs in this respect to the original Fano effect
which involves only two interfering contributions.

Some of the features of the single dot interferometer
can be understood better if we temporarily specialize Eq. (\ref{gab})
to a triangular interferometer which is exactly solvable. More precisely,
the quantum dot is a single site (thus $0a=\alpha, 0b=\beta, a=b$) and the reference arm
is reduced to the lead ends $\alpha$ and $\beta$.
Then the conductance can be presented in the generalized Fano-like form:
\begin{equation}\label{fano}
g_{\alpha\beta}(E_F)=\frac{4\tau_L^4\sin^2k}{|a^2-1|^2}\cdot\frac{|\varepsilon +q|^2}{1+\varepsilon ^2},
\end{equation}
with the following notations:
\begin{eqnarray}
\varepsilon &=&\frac{E-V_g-E_F+{\rm Re}A}{{\rm Im}A},
q=-\frac{\tau^2e^{-2\pi i\phi}+{\rm Re}A}{{\rm Im}A}\\
A &=&\frac{2\tau^2(\cos(2\pi\phi)-2)}{a^2-1},\quad
a=-\tau_L^2e^{-ik}-E_F.
\end{eqnarray}
Notice that here the Fano parameter $q$ is complex and not real as in \cite{Fano},
a fact that was suggested in the fitting conductance formula proposed in Ref.\cite{F2}
It is easily seen that $q$ is a periodic function of flux.

A complex $q$ was derived for a multichannel double barrier in \cite{Wu}
and also by Nakanishi \cite{Nak} under a double-slit condition.
 Alternative calculations for exactly solvable models were given in \cite{HKS,Ueda,KH}.
In particular, the Fano parameter obtained in Ref. \cite{Ueda} lies on an ellipse in
the complex plane with the center at origin, while according to
Kobayashi {\it et al.} \cite{F2} the center of the ellipse is shifted from the origin.
This shift is captured within our model. Indeed, by straightforward calculations
one gets explicit formulae for the real and imaginary parts of the Fano parameter \cite{ReIm}.
Clearly $q$ lies on an ellipse with the center shifted on the real axis.
The shift is due to the second term from ${\rm Re}(q)$ and disappears
only for $E_F=0$.
We believe the difference from previous calculations comes from the fact
that in our approach the leads' self-energy $\Sigma^L$
depends explicitly on energy, while in Refs.\cite{HKS,Ueda,KH} this quantity is energy independent.
The real part which is responsible for the
asymmetry of the line shape behaves roughly as $\cos2\pi \phi$
 and ${\rm Im }(q)$ as $\sin2\pi \phi$ which
agrees qualitatively with Fig.\ 6b from Ref.\cite{F2}.
The condition for a symmetric peak is achieved twice in a flux period since
${\rm Re}(q)$ vanishes twice.

\section{Results and discussion}

Following the setup
from Ref. \cite{F1} we consider a quantum dot having 7$\times$8 sites while the 1D ring
contains 140 sites. The uniform perpendicular magnetic field pierces both the ring and the QD.
Throughout the paper the magnetic flux piercing the ring will be denoted by $\phi$
and expressed in quantum flux units $\Phi_0$. Figure 2 shows several Fano line shapes of the transmittance
$g_{\alpha\beta}$ as a function of the gate potential on the dot $V_g$, for fixed magnetic flux.
The lead-ring coupling $\tau_L=1$.

Now we look for the Aharonov-Bohm oscillations of the transmittance in order to identify
specific properties due to the 2D character of the dot.
To this end we fix first an initial magnetic flux and choose two gate
potentials whose associated transmittance values are located on different sides of a Fano
peak. Then we keep the gate potentials unchanged and vary instead the flux through the ring.
The transmittances assigned to the two gate potentials (Fig.\ 3a) show AB
oscillations, proving that the transport through the system is coherent.
We notice that in the flux interval $[5,10]$ the two oscillations are not in-phase,
while for $\phi\in [10,15]$ they are in-phase. Fig.\ 3b helps us to explain this fact.
It shows that as $\phi$ increases the Fano line moves to the right (as we shall argue below,
this shift is due only to the magnetic field dependence of the resonant level of the 2D dot).
Consequently, the transmittance value corresponding to $V_g=-2.680$ passes gradually from one side
of the peak to the other. The irregularity seen in Fig.\ 3a for $\phi\in [10,12]$ is due
to the passing through the Fano peak. When the magnetic flux is chosen such that the transmittance
value for $V_g=-2.680$ is still on the right side of the peak, its Aharonov-Bohm oscillation
is in anti-phase with the one for $V_g=-2.694$. This happens because the phase of the QD transmittance
jumps by $\pi$ on resonance. If $\phi$ is further increased the resonant level of the dot moves
in magnetic field and the corresponding Fano line is pushed to the right so that the transmittance
value for $V_g=-2.680$ goes to the left side of the peak.
Then clearly the two AB oscillations are in-phase. Thus the phase of the AB oscillation can be changed not only by varying the gate potential but also by varying the magnetic flux through
the ring. This effect is important only if the dot levels depend in a sensible way on the magnetic flux.

Another effect coming from the spectral properties of the 2D quantum dot is the magnetic drift of the
Fano line, as already shown in Fig.\ 3b. In order to see this more clearly
we have monitored the position of the Fano peak (in $V_g$ units) as function of
the magnetic flux in Fig.\ 4a (the solid line). Its trajectory is rather complex:
 the global shift towards higher gate voltages is modulated by a $\Phi_0$-periodic
oscillation. The shift comes from the magnetic field dependence
of the dot levels. Indeed, the dotted line shows the trajectory of the eigenvalue of
the isolated dot which is associated to the Fano resonance located in the range $[-2.8,-2.6]$
(see Fig.\ 2). Clearly, its positive slope is responsible for the magnetic drift of the Fano
peak position. Note also that the peak positions correspond to gate voltages
$V_g\sim E_i-E_F$. This is roughly the resonant condition for the tunneling through
the dot. The dotted line represents the
trajectory of the symmetric peak that appears when the lower arm of the ring is disconnected
(i.e $\tau=0$). In this case there is no interference, the Peierls phases disappear from
Eq. (\ref{gab}) and the only flux dependence comes from the denominator of the effective
Green function. Because of this the symmetric peak position has a simple drift. The above
discussion shows also that the additional modulation of the Fano peak position
is due entirely to the other flux dependences in Eq. (\ref{gab}).

A parameter which provides criteria for the experimental observation of the two magnetic
field effects reported above is the ratio $R$ between the area of the ring $A_r$ and
the area of the dot $A_d$. For the system composed of a $7\times8$ sites dot and an
140 sites ring $R=20$ while in the experiments of Kobayashi {\it et al.} $R\sim 56$.
Fig.\ 4b shows the behavior of the Fano peak position for $R\sim 44$ and $R\sim 70$.
We kept the dot dimension fixed and increased the length of the ring; for $R\sim 44$
the ring has 200 sites and for $R\sim 70$ the ring has 250 sites.
The data from Fig.\ 4a were added in Fig.\ 4b in order to make the comparison easier.
In the case of a 250 sites ring the Fermi level was set to $E_F=0.02$ while in
the other two cases $E_F=0.05$ (this causes the large shift of the peak position for
the 250 sites ring). There are two things to be noticed: i) as the ring gets bigger
the oscillation of the peak position becomes simpler - the two local maxima located on
each side of the absolute maxima for $R\sim 20$ are pushed towards integer flux values
for $R\sim 44$ and vanish completely at $R\sim 70$. ii)
The magnetic drift of the peak position is weaker for $R\sim 44$
(one can notice it only at rather large fluxes) and disappears for $R\sim 70$.
This is easily understood since for bigger ring a flux quanta means smaller magnetic
fields inside the dot, thus a slower deviation of its eigenvalues. Having in mind
the ratio $R\sim 56$ from Ref. \cite{F1} it is clear that
the oscillations like the ones depicted in Fig.\ 4a and the magnetic drift of
the Fano lines were not observed due to the large ratio $R$.

In the following we discuss the magnetic field dependence of the Fano interference.
We look first at the Fano dip position in magnetic field.
The solid curve in Fig.\, 5 gives the dip trajectory. As already
explained for the simple triangular interferometer the Fano peak has two domains
of symmetry between consecutive integer flux values.
In this case the Fano dip cannot be defined and the eye-guiding dashed line marks the
sudden change of the dip position from one side of the resonance to the other.

The next step of our analysis is to study the phases of the two complex terms
contributing in Eq. (\ref{gab}). We shall denote by $\varphi_{bg}$ the phase
  of the background contribution ${\tilde G}^R_{\alpha\beta}$ while
$\varphi_{res}$ stands for the phase of the second term. We stress that
$\varphi_{res}$ cannot be simply identified with the intrinsic phase of
the quantum dot since it is the phase accumulated along multiple bouncing
trajectories in the inteferometer.
Fig.\, 6a shows the evolution of a Fano line shape from Fig. 2 for other two values of
the magnetic flux. The asymmetric tail changes periodically with the magnetic flux as
reported in Ref. \cite{F1}. More precisely, the Fano dip located on the right side of
a transmittance peak for $\phi=5.0$ moves to the other side as the flux is increased by a
half flux quanta and comes back on the RHS when $\phi=6.0$. The filled circles mark the
transmittance values at $V_g=-2.9616$. This value was chosen simply because, at $\phi=5.0$,
it corresponds to a Fano peak. Fig.\ 6b depicts the phases $\varphi_{bg}$, $\varphi_{res}$ and
their difference $\Delta\phi=\varphi_{bg}-\varphi_{res}$
as a function of flux while keeping $V_g=-2.9616$. The phase $\varphi_{bg}$ is simply a
linear function of flux, decreasing by $\pi$ as one more flux quanta pierces the ring.
$\varphi_{res}$ presents instead a weak oscillation and increases by $\pi$, such that the
phase difference decreases roughly by $2\pi$. The investigation of Fig.\ 6b allows us to
explain the transition from the Fano peak at $\phi=5.0$ to a Fano valley at  $\phi=5.5$
and finally the recovery of the constructive interference at $\phi=6.0$. It is easy to see
that the modulus of ${\tilde G}^R_{\alpha\beta}$ in Eq. (\ref{gab}) does not depend on
the magnetic field thus it is a constant complex vector which rotates clockwise according to
its flux-dependent phase $\varphi_{res}$.
The vector describing the second term in Eq. (\ref{gab}) depends on $\phi$ both in
amplitude and phase and rotates anti-clockwise (not shown).
At $\phi=5.0$,  ${\tilde G}^R_{\alpha\beta}$ lies almost on the
positive real axis while the resonant term is in the 4-th quadrant, its phase being close to
$\pi/2$. The real parts of the two contributions are both positive and the imaginary
ones negative,
thus they interfere constructively justifying the Fano peak at $\phi=5.0$. In contrast, for $\phi=5.5$
the background vector lies on the negative imaginary axis while the resonant one comes close to the
positive imaginary axis. Clearly the interference is destructive and the transmittance assigned to
$V_g=-2.6916$, $\phi=5.5$, is located in a Fano valley. Using similar arguments one can
explain the situation from $\phi=6.0$. We have checked that the results mentioned above
hold as well for all other Fano resonances from Fig.\ 2.

\section{Conclusions}

We have studied the mesoscopic Fano effect in an Aharonov-Bohm ring with a two-dimensional
quantum dot. We have considered the effect of the magnetic field on the orbital motion of
electrons inside the dot. It is shown that this dependence leads to a global shift
of the Fano lines as a function of the magnetic field.
It would be interesting to perform experiments with bigger dots in order to observe this effect.
However, if the dot area is small compared to the ring area as in Ref. \cite{F1}
or if the magnetic field is weak this shift can be neglected.
We emphasized that a continuous variation of the AB oscillation phase by
$\pi$ along one Fano resonance can be also obtained by varying the magnetic flux and
keeping the gate potential fixed. The origin of this effect lies again in the
specific spectral properties of the two-dimensional dot.

The mechanism leading to the magnetic control of the
Fano interference reported in the experiments of Kobayashi {\it et al.} \cite{F1}
was described by a detailed analysis of the phases of two complex contributions appearing in the
conductance formula: the reference transmittance of the free arm and {\it all} possible
tunneling processes through the dot.

\acknowledgments{
V.\,M. was supported by a NATO Science Fellowship.
We acknowledge instructive discussions with Prof. B. R. Bu{\l}ka. }


\begin{figure}
\includegraphics{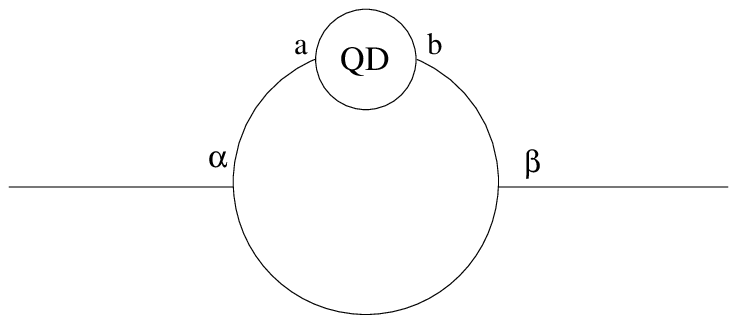}
\caption{The sketch of the AB interferometer. $a,b$ are the sites
of the dot that are coupled to the ring. } \label{figure1}
\end{figure}

\begin{figure}
\includegraphics{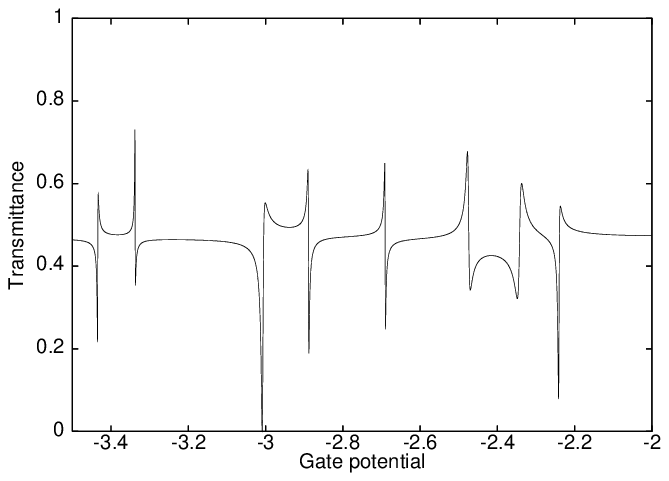}
\caption{Fano line shapes of the interferometer transmittance as a
function of the gate potential $V_g$ at $\phi=5.0$. The ring-dot
coupling $\tau=0.4$ and the Fermi level $E_F=0.05$. }
\label{figure2}
\end{figure}

\begin{figure}
\includegraphics{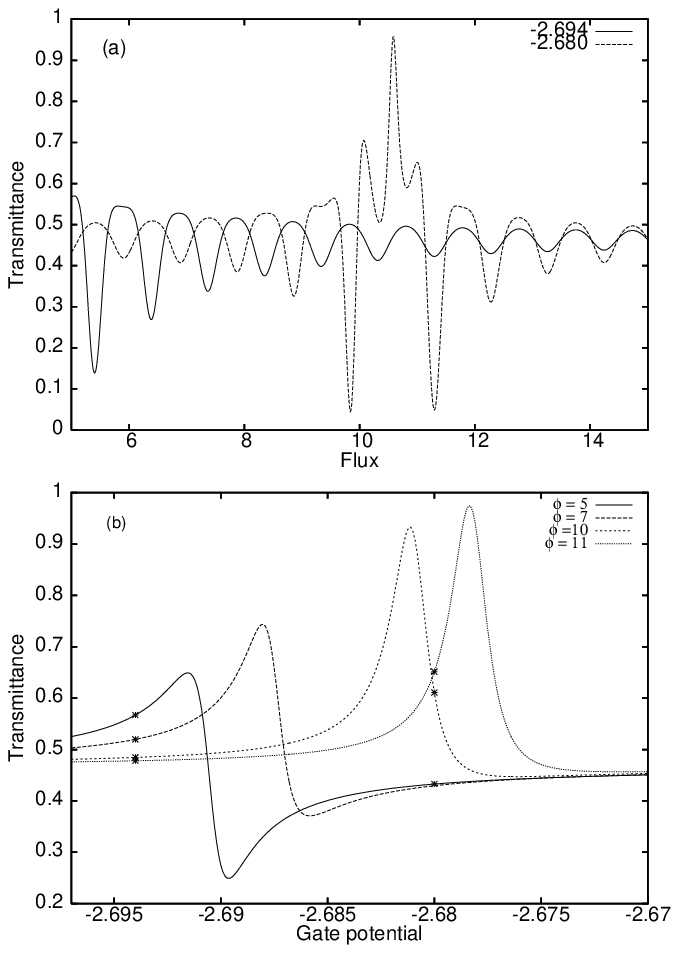}
\vskip -5cm \caption{a) The Aharonov-Bohm oscillations of the
transmittance assigned to two gate voltages $V_{g1}=-2.695$ and
$V_{g2}=-2.68$. At $\phi=5$ the two gate potentials stay on
opposite sides of the Fano line shape. b) The 'drift' of the Fano
peaks in magnetic field. The points on the line shape corresponds
to  $V_{g1}$ and $V_{g2}$ respectively. $\tau=0.4$, $E_F=0.05$. }
\label{figure3}
\end{figure}

\begin{figure}
\includegraphics{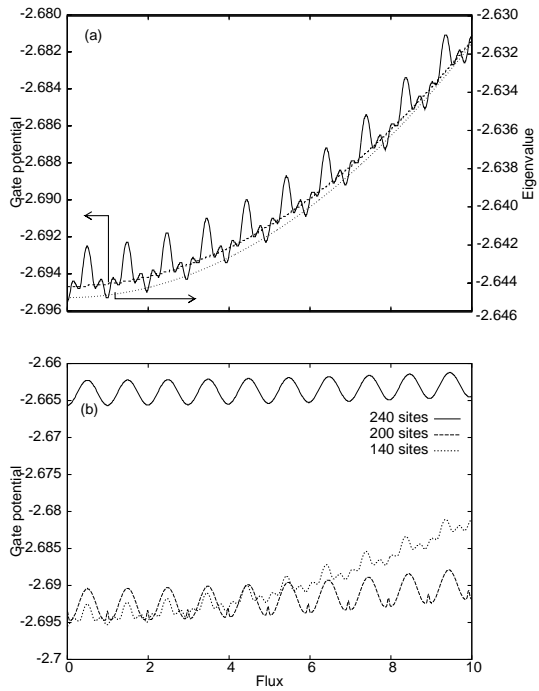}
\caption{a) The quasiperiodic trajectory of the Fano peak position
as a function of the magnetic flux. Its maximal deviations from
the initial positions along a one flux quanta variation correspond
to half-integer flux values $\tau=0.4$, $E_F=0.05$. The general
drift of the oscillation is due to the magnetic field dependence
of the isolated dot level associated to the chosen Fano peak
(dotted line - its corresponding axis is on the right). The nearby
dashed line represents the position of the symmetric transmittance
peak which appears when the lower arm is disconnected (i.e
$\tau=0.0$). (b) The effect of the ring size on the peak position
oscillations. } \label{figure4}
\end{figure}

\begin{figure}
\includegraphics{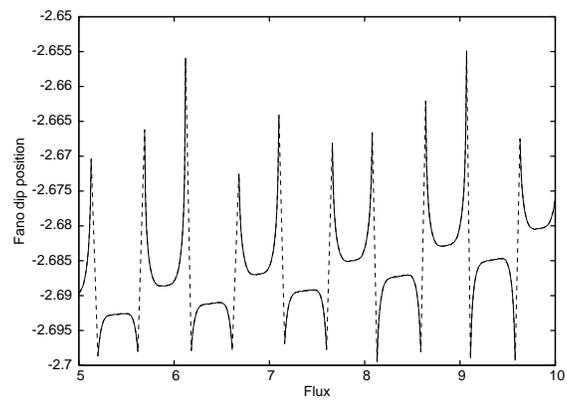}
\caption{The position of the Fano dip in the flux domains where it
can be defined (solid line). The dashed line underlines the
switching of the dip from one side of the resonance to the other.
Note that inside each interval $(N\Phi_0, N\Phi_0+1)$ there exists
a regime where the dip position is almost constant.}
\label{figure5}
\end{figure}

\begin{figure}

\includegraphics{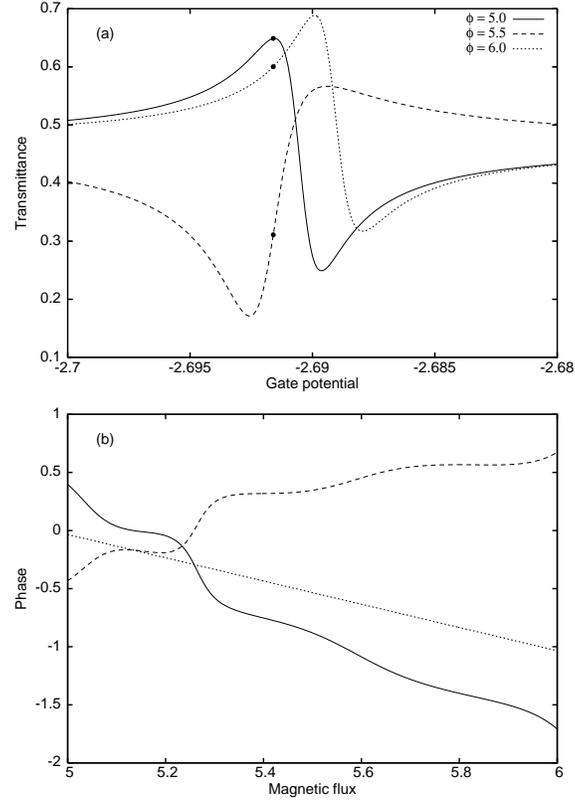}
\vskip -10cm \caption{a) The magnetic control of the Fano
resonance for $V_g=-2.6916$: at $\phi=5.0$ and $\phi=6.0$ the
interference is constructive while at $\phi=5.5$ is destructive.
b) The phases of the two terms in Eq. (\ref{gab}) ($\phi_{ bg}$ -
dotted line, $\phi_{ res}$ - dashed line) and the phase difference
$\Delta\phi$ (solid line) as functions of magnetic field.
$\tau=0.4$, $E_F=0.05$.} \label{figure6}
\end{figure}

\end{document}